# What to do if N is two?

Pascal Fries[1,2] & Eric Maris[2]

[1] Ernst Strüngmann Institute (ESI) for Neuroscience in Cooperation with Max Planck Society, Deutschordenstraße 46, 60528 Frankfurt, Germany.

[2] Donders Institute for Brain, Cognition and Behaviour, Radboud University Nijmegen, Montessorilaan 3, 6525 EN Nijmegen, Netherlands.

## Summary

The field of in-vivo neurophysiology currently uses statistical standards that are based on tradition rather than formal analysis. Typically, data from two (or few) animals are pooled for one statistical test, or a significant test in a first animal is replicated in one (or few) further animals. The use of more than one animal is widely believed to allow an inference on the population. Here, we explain that a useful inference on the population would require larger numbers and a different statistical approach. The field should consider to perform studies at that standard, potentially through coordinated multi-center efforts, for selected questions of exceptional importance. Yet, for many questions, this is ethically and/or economically not justifiable. We explain why in those studies with two (or few) animals, any useful inference is limited to the sample of investigated animals, irrespective of whether it is based on few animals, two animals or a single animal.

## Introduction

In awake monkey neurophysiology, there is the saying: "One monkey is an anecdote, two monkeys are perpetual truth." This captures the current publication standard in the field of non-human primate research. Yet, the situation is not unique to this field. Rather, in-vivo neurophysiology studies generally and also parts of psychophysics tend to report data from very few participants per study. In fact, studies with a low number of participants appear in every field where the number of participants is restricted by either ethical or economical considerations. Such studies were also common in the early days of human brain mapping, when this field relied heavily on the use of radiotracers for positron emission tomography (PET). Partly because functional magnetic resonance imaging (fMRI) became widely available, the field considered the role of sample size several decades ago, made a fundamental distinction between random- and fixed-effects analyses, and developed the notion of conjunction analysis. We aim to open up these fundamental statistical concepts to any field of neuroscience that typically uses two (or few) animals or participants. We do not aim here to present novel or even optimal statistical techniques. Rather, we focus on standard statistical approaches and explain what can and what cannot be inferred from them when the sample size is two or few. The insights are particularly relevant for any study using animals, because the number of used animals ought to be reduced as far as possible for ethical reasons. We will therefore mainly refer to (the number of) animals, yet the arguments apply equally to any field of science that bases statistical tests on few participants. In the following, we take "few" to mean "less than five", although this exact number does not play a crucial role in our arguments.

Statistical tests are often reported for each of few animals, and the qualitative similarity between animals is mentioned, or the data from few animals are pooled. In both cases, the fact that more than one animal is involved is usually considered crucial. The rationale for requesting at least a second animal is that it provides a replication. And the rationale for a replication is that this allows further-reaching inferences. Any scientific study on animals ideally aims for an inference on the population from which these animals are sampled, rather than an inference that is limited to the studied sample of animals. But does a replication in two to few animals permit this population-level inference? The main messages of this paper are 1.) that replication across few animals does not allow for a formally well-founded and useful inference on the population of animals; 2.) that such an inference would require more animals to gain acceptable sensitivity; 3.) that the simple combination of data across animals (i.e., treating the data as if they were generated by a single animal) is often the best one can do; 4) that the resulting inference is limited to the studied sample and therefore does not qualitatively differ between few, two or even a single animal.

## Fixed-effect versus random-effect tests

The argument starts with the basic rationale by which statistical tests draw inferences. This rationale leads us to the important difference between fixed- and random-effect tests, which is directly relevant to whether one can infer on a population of animals. Let us consider a study that has recorded, in one animal, the firing rates of 100 neurons, each in two experimental conditions. The study compares the two conditions and finds an effect: the average firing rate (with averaging over the neurons) is higher in condition B than A, and a properly controlled statistical test gives a p-value of less than 0.05. This allows the following inference: this observation on 100 neurons would have a probability of less than 0.05, if the average effect calculated on all neurons that could be recorded from this animal would be zero, i.e. if the null hypothesis were true. Therefore, we make the inference that the effect is most likely present, when calculated on all neurons.

The same would hold if the unit of observation (UO, the entity to which the observations pertain; (Maris, 2012)) is not the neuron, but e.g. the trial. Let us consider a study that uses optogenetics to stimulate a certain brain area in one animal. The study compares reaction times (RTs) in a detection task between a condition with and one without optogenetic stimulation. For both conditions, 200 trials are obtained, and RTs are compared across trials. In this case, the inference is on all trials that could be obtained in the same way in this one animal. Along the same line, if RTs are averaged per session and statistics are performed across sessions, the inference is on all sessions that could be obtained in the same way in this one animal.

Crucially, the inference is always on a population of neurons (or trials or sessions) and for this (the type of elements in this population), it does not matter whether they are pooled over one, two or few animals. If neurons are obtained from one animal, the inference is on the population of neurons from that specific animal; and if data are pooled over e.g. three animals, the inference is on the population of neurons from those specific three animals. In this sense, the inference remains identical, irrespective of whether few animals, two animals or even a single animal are used. Crucially, when pooling over multiple animals, the inference does not pertain to a population of animals.

For example, if 60 neurons were recorded from animal 1 and 40 from animal 2, a common approach would be to pool neurons over animals, and to perform a single t-test across these 100 neurons. The population to which one infers on the basis of this single t-test is a population of neurons of which 60 percent is obtained from animal 1, and 40 percent from animal 2. As in the situation with a single animal, the inference is to a population of neurons, and not to a population of animals.

Of course, it is also possible to report separate tests for the individual animals, and this allows for a different type of analysis, which we will treat in the section *Conjunction analysis as a formal replication approach*.

We now consider inference to a population of animals, which requires that data are obtained from multiple animals. Specifically, one can then average the relevant quantities (like spike rate) **within** each animal (over neurons, trials, sessions), and perform a statistical test **across** animals. In this statistical test, the animals have the same role as the neurons, trials or sessions in the statistical tests that we considered previously. In all cases (neurons, trials, sessions and animals), the same mathematical logic applies. But there is a crucial step when moving from a sample obtained **within** one or a few animals to a sample **of** animals. Testing using a sample of animals allows to make an inference on the population of animals. Testing using a sample of neurons, trials or sessions sampled from one or a few animals merely allows an inference on **these** animals, and this is independent of whether sampling is done over few animals, two animals or a single animal.

If the UOs are neurons, trials, or sessions, an effect is typically called a **fixed effect**, because the inference is bound to the fixed sample of animals from which the UOs were sampled. If the UOs are animals, an effect is typically called a **random effect**, because the inference pertains to some population of animals from which a sample was randomly drawn.

## A random-effect test requires many animals to reach significance

Given that most in-vivo neurophysiological studies involve more than one animal, a random-effect statistical test can in principle be performed across this small number of animals. Unfortunately, a random-effect statistical test using the data of few animals typically has a very low sensitivity (statistical power, the probability of rejecting the null hypothesis when it is in fact false). To understand why this is the case, it is useful to investigate the factors that determine the outcome of a statistical test. For concreteness, we will now focus on the paired- or dependent-samples t-test, although our argument is equally valid for many other tests. The paired-samples t-statistic (denoted by $T$) depends on three factors: (1) the average between-condition difference, or effect size, ($M$) in the sample, (2) the standard deviation ($s$) across the sample of UOs of the effect sizes (i.e. effect sizes calculated per UO), (3) the number of UOs ($N$). Formally,

$$T = \frac{\sqrt{N} \times M}{s}$$

For a typical in-vivo neurophysiology study, $N$ is fixed to a small number, such that $T$ depends solely on the ratio $M/s$, also called "Cohen's d in the sample". Even if $M$ is large, $T$ will be insignificant if $s$ is too large. Unfortunately, when $N$ is small, only a very narrow range of raw data values results in $s$-values that are small enough for observing a significant $T$. Consider a study comparing two conditions in each of two animals (and testing a two-sided hypothesis, i.e. making no assumption about the sign of the effect). In condition A, both animals show an average firing rate of 10 Hz. In condition B, the first animal shows a firing rate of 20 Hz, i.e. a 100% increase. For $T$ to be significant at the 0.05 level, the second animal needs to show an increase between 85% and 115%. Larger deviations *in either direction* would increase $s$ so much that $T$ would become insignificant.

The minimum number of animals for a random-effect test can be determined if 1) we assume that there is an effect with an effect size given as Cohen's d in the population, and 2) we specify a desired sensitivity (typically 0.8). Then, the minimum number of animals follows from standard calculations (Cohen, 1992). For a sample Cohen's d of one, the t-test would only reach significance for a sample of at least five animals. A similar reasoning applies if only the signs of the effects of the individual animals are relevant, in which case we can use the sign test. A sign test can only be significant (p<0.05) with at least five animals for a one-sided, and with at least six animals for a two-sided hypothesis.

## Conjunction analysis as a formal replication approach

A random-effect statistical test pertains to an average effect in some population. An alternative way to make an inference about a population starts from the intuition that multiple individually significant fixed-effect tests provide more evidence than a single significant fixed-effect test, corresponding to the current tradition of in-vivo neurophysiology: A significant test in a first animal is often replicated in a second (or few) animals, and if the test is significant in each of them, this is taken as evidence for the effect being typical in the population of animals. This intuition has been turned into a formal statistical method, which is referred to as conjunction analysis (Friston et al., 1999).

A conjunction analysis takes an individually significant fixed-effect test as evidence for the presence of an effect in that participant. A conjunction analysis only makes sense if the direction of this effect (A>B or B>A) is the same for all participants and specified a priori. Given the number of participants that each show the effect, conjunction analysis estimates the expected proportion of the population showing the effect, which is called *typicality* $\gamma$. Specifically, the estimated typicality depends on the following parameters: 1) the number of participants $N$ that each show the effect; 2) the false-positive rate $\alpha$ of the individual fixed-effect tests; 3) the true-positive rate (sensitivity) $\beta$ of the individual fixed-effects tests.

For explaining how the typicality $\gamma$ is estimated, we start from the situation in which all individual fixed-effect tests in a sample of $N$ animals are significant; the procedure is similar for other numbers of significant fixed-effect tests. Now, the probability that all individual fixed-effect tests are significant is the following:

$$P(\text{all } N \text{ tests significant}) = [\alpha \times (1 - \gamma) + \beta \times \gamma]^N$$

Further mathematical considerations show that we cannot estimate $\gamma$ directly. Yet, we can estimate a useful lower bound to $\gamma$, called $\gamma_c$, reflecting the minimal proportion of the population for which we can expect the effect to be present. For this, we first specify $P(\text{all } N \text{ tests significant}) = 0.05$, and use essentially the same rationale as the one with which the false-positive rate is controlled: if the effect is absent (rare) in the population, there should be a probability of less than 5% for the test to reach significance. Thus, we can rewrite the formula as follows:

$$0.05 = [\alpha + (\beta - \alpha)\gamma_c]^N$$

$\alpha$ (the false-positive rate of the fixed-effect tests) is typically set to 0.05. Thus, a small proportion of significant tests are false positives, slightly reducing the estimated typicality. $\beta$ (the true-positive rate of the fixed-effect tests) cannot be determined, because it is the proportion of significant fixed-effect tests that are truly positive, and we have no access to this truth. If we would assume that sensitivity was less than one, a non-significant fixed-effect test could reflect a false-negative test (and therefore essentially be discarded), allowing for any typicality to be consistent with the outcomes of the fixed-effect tests. This can be concluded directly from the formula: If we chose $\beta$ to be less than one, the resulting lower bound $\gamma_c$ would increase without limit, which is obviously meaningless. Therefore, we need to make the conservative assumption that sensitivity is one, as has been argued before (Friston et al., 1999).

Using $\alpha = 0.05$ and $\beta = 1$, and numerical methods to solve the formula, Figure 1 shows $N$, the number of participants, as a function of $\gamma_c$, the lower bound to typicality. This allows several important insights: 1.) With an individually significant effect in two participants, one can merely infer that the probability to find the effect in the population is at least 0.18; for three participants, this value increases to 0.34. 2.) We propose that 0.5 is the lowest useful value for typicality, because it corresponds to the expected presence of an effect in at least a simple majority of the population. For a typicality of 0.5, the number of required participants is five. 3.) For larger desired values of typicality, the number of required participants rises steeply: a typicality of 0.7 requires 9 participants, and a typicality of 0.9 requires 30 participants.

It is important to note that the required number of participants hardly changes if we use a lower nominal false-positive rate $\alpha$ for the individual fixed-effect tests (which is under the researcher's control) and keep $P(\text{all } N \text{ tests significant})$ at 0.05. The number strongly increases, if we lower both $\alpha$ and $P(\text{all } N \text{ tests significant})$ by the same amount.

Conjunction analysis provides a formal evaluation of the inferences that can be based on replication across few animals. It reveals that a replication in a second or third animal does enable an inference on the population that goes beyond the inference of a single-animal fixed-effect analysis. However, it also reveals that this additional inference is very limited. Scientists would have to agree that there is added value in knowing that there is a 0.18 or 0.34 probability of finding the effect in the population, because this is the inference that can actually be drawn from replication in one, respectively two, additional animals. This inference falls far short of the assumption that a second or third animal allows to make an inference on the population. The responsible scientist will need to take this into account, when deciding on the number of animals used in a study. Notably, the formal analysis shows that it is highly questionable when reviewers require an individually significant effect in a second or third animal, with the explicit or implicit argument that this is the international standard.

## Conclusions and recommendations

Our analysis leads to the following conclusions:

- There are no formal arguments against a fixed-effect statistical test on the combined data of all animals. However, crucially, scientists should be aware that such a test does not warrant an inference on the population of animals. The inference through a fixed-effect test remains limited to the studied sample of animals, irrespective of whether the conclusions are based on a single animal, two animals, few animals or even many animals.
- Few animals, in fact as few as two animals, are mathematically sufficient to perform a random-effect test across those animals. However, this test would suffer from extremely low sensitivity, i.e. an extremely high false-negative rate.
- A conjunction analysis provides a formal evaluation of the inferences that can be based on replication across few animals. It reveals that a replication in few animals does enable an inference on the population that goes beyond the inference of a single-animal fixed-effect analysis. However, it also reveals that this additional inference is too limited to be of interest for the scientific community. Therefore, the tradition to provide (or request) individually significant results in each of few animals is problematic. If not combined with an explicit mentioning of its severe limitations, i.e. the obtained low typicality, it likely misleads many readers.

Based on these conclusions, we make the following recommendations for in-vivo neurophysiology:

- If there ***is not*** a pre-existing dataset:
    - Investigators should first obtain and analyze data from a single animal and publish them. Editors and reviewers, who so far routinely accepted two-animal papers (or few-animal papers), should routinely accept single-animal papers, because the inference is in both cases on the sample and not on the population of animals.
    - For most respective studies, this new standard will reduce the number of animals by at least 50%, which is a substantial contribution to animal welfare.
    - A minority of studies will require multiple animals to obtain sufficient data for acceptable sensitivity, even for an inference on the sample of animals, e.g. studies that intracellularly record and label one neuron per animal.
    - After such a study with inference on the sample of animals, investigators can decide about a subsequent study, testing for an inference on the population of animals.
        - A random-effect test allows an inference on the average effect in the population, i.e. the most common goal. In most cases, this will require at least five animals.
        - A conjunction analysis allows the estimation of typicality. We propose that the minimal useful typicality is 0.5, and the corresponding required number of animals is five.

- If there ***is*** a pre-existing dataset:
    - If there are data from more than five (or six, for a two-sided sign test) animals, investigators should consider a random-effect test, for inferring the average effect, or a conjunction analysis, for inferring typicality.
    - If there are data from less than five animals, a fixed-effect test on the pooled data from all animals should be performed and reported. Reviewers and editors should not require tests per animal, or even significance per animal.

In brief, provided one carefully qualifies the nature of one's inference, fixed-effect analyses of data from a single animal are perfectly valid to establish an existence proof for the effect in question. Providing this existence proof is orthogonal to the assertion that this feature or effect is conserved over a population. In many instances, random-effects analyses (and related conjunction analyses) may not be necessary. The canonical example here is the 'talking dog'. If you were to decide about publication, what N would you require before accepting a report of talking dogs?

Figure legend:

Figure 1: The number of participants required to infer a given typicality.

The main figure and the inset show the number of required participants as a function of the level of required typicality. The inset provides an enlarged version of the parameter range discussed in the text. The blue lines give the real-valued numbers calculated according to the formula in the text. As only whole participants can be added to the sample, the actually necessary integer numbers of participants are shown as red lines.

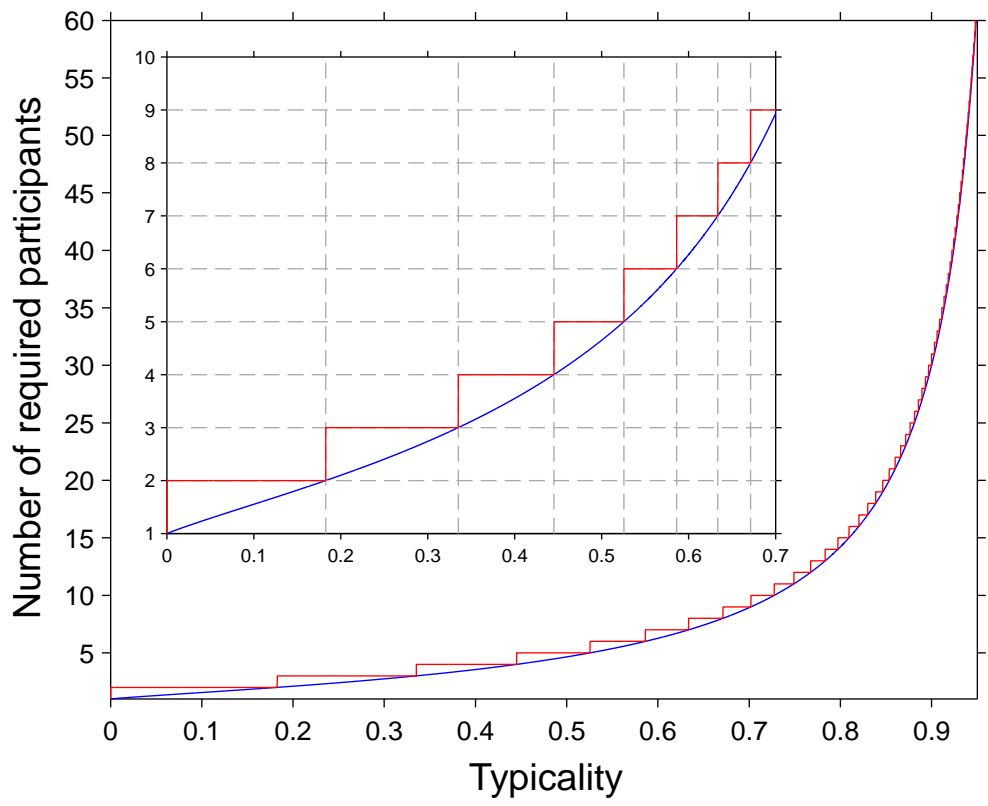

Fries and Maris, Figure 1